\documentclass[article,5p,twocolumn]{elsarticle}
\usepackage{graphicx,latexsym}
\usepackage{amssymb,amsmath,bm}
\usepackage{subfigure}
\usepackage{braket}

\usepackage{hyperref}
\hypersetup{
    pdfnewwindow=true,       % links in new window
    colorlinks=true,         % false: boxed links; true: colored links
    linkcolor=blue,          % color of internal links
    citecolor=blue,          % color of links to bibliography
    filecolor=magenta,       % color of file links
    urlcolor=black           % color of external links
}

\def\eq#1{Eq.\ (\ref{#1})}
\def\mb#1{\mbox{\boldmath$#1$}}
\def\fig#1{Fig.\ \ref{#1}}

\begin{document}

\begin{frontmatter}

%-----------------------------------------------------------------
\title{Effects of photon field on heat transport through a quantum wire attached to leads}

\author[a1,a2,a3]{Nzar Rauf Abdullah}
\ead{nzar.r.abdullah@gmail.com}
\address[a1]{Physics Department, College of Science, University of Sulaimani, Kurdistan Region, Iraq}
\address[a2]{Komar Research Center, Komar University of Science and Technology, Sulaimani City, Iraq}
\address[a3]{Science Institute, University of Iceland, Dunhaga 3,
        IS-107 Reykjavik, Iceland}

\author[a4]{Chi-Shung Tang}
\address[a4]{Department of Mechanical Engineering,
        National United University,
        1, Lienda, Miaoli 36003, Taiwan}
        
\author[a5]{Andrei Manolescu}
\address[a5]{School of Science and Engineering, Reykjavik University,
        Menntavegur 1, IS-101 Reykjavik, Iceland}
        
\author[a3]{Vidar Gudmundsson}

%
%----------------------------------------------------------------

\begin{abstract}
We theoretically investigate photo-thermoelectric transport through a quantum wire in a photon cavity 
coupled to electron reservoirs with different temperatures. Our approach, based on a quantum master equation, 
allows us to investigate the influence of a quantized photon field on the heat current and thermoelectric transport in the system. 
We find that the heat current through the quantum wire is influenced by the photon field resulting in a negative heat current
in certain cases.
The characteristics of the transport are studied by tuning 
the ratio, $\hbar\omega_{\gamma} / k_{\rm B} \Delta T$, between the photon energy, $\hbar\omega_{\gamma}$, 
and the thermal energy, $k_{\rm B} \Delta T$. 
The thermoelectric transport is enhanced by the cavity photons when $k_{\rm B} \Delta T > \hbar\omega_{\gamma}$.
By contrast, if $k_{\rm B} \Delta T < \hbar\omega_{\gamma}$, the photon field is dominant and a suppression 
in the thermoelectric transport can be found in the case 
when the cavity-photon field is close to a resonance with the two lowest one-electron states in the 
system. Our approach points to a new technique to amplify thermoelectric current in nano-devices.
\end{abstract}


\end{frontmatter}

\

\section{Introduction}

The wide research field of thermoelectric transport in nanoscale devices is very active nowadays due to 
their expected high efficiency comparing to bulk materials.
The efficiency of thermoelectric materials is measured by the figure of merit, 
the ratio of the electrical conductance 
to the thermal conductance~\cite{PhysRevB.82.235428}. 
In bulk materials, the figure of merit is restricted by classical relationships such as the Wiedmann-Franz law and 
the Motto relation in which the electrical conductance is directly proportional to the thermal conductance.
The violation of the Wiedmann-Franz law in the range of the nano-scale has caused nanostructures 
to be considered as good thermoelectric devices~\cite{PhysRevB.86.035433}. These relations may not hold in 
nanostructures due to quantum phenomena such as quantum interference~\cite{PhysRevB.84.075410}, 
Coulomb blockade~\cite{PhysicaE.53.178}, and energy quantization~\cite{Majumdar303.777}.

It has been shown that the thermoelectric efficiency in nanodevices is increased~\cite{PhysRevB.85.085408} 
and plateaus in the thermoelectric current due to Coulomb blockade
can be formed~\cite{PhysicaE.53.178} in the presence of the Coulomb interaction. 
In addition, the thermal properties of a double quantum dot molecular junction have been studied and 
shown that the Fano effect can improve the thermoelectric efficiency~\cite{Liu.108.2010}.
Thermoelectric effects in an Aharonov-Bohm interferometer with embedded quantum dots has been demonstrated~\cite{PhysRevB.67.165313}, 
where the effects of a geometrical phase on the thermopower in the absence of an interdot Coulomb interaction is shown.
On other hand, the photon-phonon interaction also influences thermoelectric transport in quantum systems. 
Soleimani~\cite{Tagani201336} has shown that the photon-phonon interaction can enhance the thermoelectric effect 
in molecular devices, with increasing oscillations of the thermopower in the presence of the 
electron-photon interaction.

A temperature gradient can also generate thermo-spin transport.
The thermo-spin effect opens up a new possibility for fabricating spintronic devices. 
It has been shown that the thermo-spin effect in nanodevices is conceptually different from the traditional 
one in bulk material using magnetic semiconductors~\cite{Sinova.880.2010}.
With the investigation of the thermo-spin effect, intensive study has been carried out on the 
thermal properties in quantum dots with a Rashba-spin orbit interaction~\cite{Zhu20104269}.

Another interesting aspect is the use of light to control thermoelectric transport in quantum systems. 
The thermoelectric effect in a quantum dot in the presence of microwave field 
has been studied using Keldysh nonequilibrium linear-response approach. 
It was found that the microwave field can produce heat flow in the QD system by a formation of 
additional transport channels below the Fermi energy~\cite{Feng.24.145301}. 
Furthermore, the thermoelectric transport properties
of topological insulator have been investigated under the application of off-resonant
light. It is shown that by applying a circularly polarized
light, the band gap is tuned and results in enhanced thermoelectric
transport. Moreover, an exchange of the conduction
and valence bands of the symmetric and antisymmetric
surface states is seen by tuning the light polarization~\cite{PhysRevB.91.115311}.

The thermo-spin transport has been theoretically studied with the application of circularly polarized light in two regimes:
Low and high temperature. At high temperature, the transport of the spin-down electron plays a dominant role 
in driving the thermoelectric effect. Alternatively, 
at low temperature, the polarized light induces an antiresonant transport of spin-up electron through the device, 
leading to an important contribution of the spin-up electrons to the thermoelectric effect~\cite{doi:10.7566/JPSJ.82.014603}.

The influences of light on the thermoelectric effect in quantum systems is still in its infancy.
Especially, if the light is quantized such as photons in a cavity. In this study, we
present theoretical investigation of thermoelectric transport through a quantum wire coupled to a quantized photon cavity.
We study two different regimes: Low and high thermal energy comparing to the photon energy. 
In the high thermal energy regime ($\hbar\omega_{\gamma} < k_{\rm B} \Delta T$), the temperature gradient is dominant, 
and the thermoelectric transport is enhanced. 
But in the low thermal energy regime ($\hbar\omega_{\gamma} > k_{\rm B} \Delta T$), 
a reduction in the thermoelectric transport is found for the system close to a Rabi resonance.

The outline of the paper is as follows. We show the model system and discuss the quantum master 
equation in Sec.~\ref{Method_and_Theory}.
Numerical results are discussed for the model in Sec.~\ref{Results}. 
Finally, conclusions are drawn in Sec.~\ref{ConclusionS}.

\section{Theory}\label{Method_and_Theory}

In this section, we introduce the Hamiltonian of the system 
and the formalism that defines the properties of the electron transport.
We assume a quantum wire (QW) connected to two electron reservoirs (leads) and exposed to
a quantized cavity photon field. The QW is coupled to a lead with high temperature on the left side and 
another lead with lower temperature on the right side.
The hot (h) and cold (c) leads each obey a Fermi-Dirac contribution 
\begin{equation}\label{Eq_1}
 f_{h/c} = \Big[ 1 + \exp{\big((E-\mu_{h/c})/(k_{\rm B} T_{h/c}) \big)} \Big]^{-1},
\end{equation}
where $\mu_{\rm h} (T_{\rm h})$ and $\mu_{\rm c} (T_{\rm c})$ are the chemical potential (temperature) 
of the hot and cold leads, respectively. We consider the voltage difference 
$eV_{\rm bias} = \mu_{\rm h} - \mu_{\rm c}$ to be applied 
symmetrically across the device such that the electrochemical potential of the hot and 
the cold leads is equal.

One can calculate the heat current (${\rm HC}$) that is the rate at which heat 
is transferred over time. In our system the heat current in our system can be expressed as

%\begin{align}\label{Eq_2_0_!!}
% {\rm HC} & =  \frac{d}{dt} \Big[\hat{\rho}_S(t) (\hat{H}_S-\mu \hat{N}_{\rm e})\Big] \nonumber \\
%            & = \sum_{\alpha \beta} (\hat{\alpha} | \dot{\hat{\rho}}_S | \hat{\beta})  (E_{\alpha} - \mu \hat{N}_{\rm e}) \delta_{\alpha \beta},
%\end{align}

\begin{align}\label{Eq_2_0}
 {\rm HC} & =  \mathrm{Tr}\Big[\dot{\hat{\rho}}_S(t) (\hat{H}_S-\mu \hat{N}_{\rm e})\Big] \nonumber \\
            & = \sum_{\alpha \beta} (\hat{\alpha} | \dot{\hat{\rho}}_S | \hat{\beta})  (E_{\alpha} - \mu \hat{N}_{\rm e}) \delta_{\alpha \beta},
\end{align}
where $\hat{\rho}_S$ is the reduced density operator, $\hat{H}_S$ is the Hamiltonian of the electrons in the central system 
coupled to the photons in the cavity, $\hat{N}_{\rm e}$ is the electron number operator, and $\mu = \mu_{\rm h} = \mu_{\rm c}$.
But, the rate at which electrons are transferred over time by a temperature gradient is the thermoelectric current (TEC).
The TEC through the QW connected to the leads and coupled to 
the photon cavity is defined as
\begin{equation}\label{Eq_2}
 {\rm TEC} =: \mathrm{Tr}[\dot{\hat{\rho}}_S^h(t) \hat{Q}] - \mathrm{Tr}[\dot{\hat{\rho}}_S^c(t) \hat{Q}] .
\end{equation}
Herein, the first (second) term of \eq{Eq_2} is the current from the hot (cold) reservoirs to the QW, respectively, 
and $\hat{Q} = e \hat{N}$ is the charge operator with the electron number operator $\hat{N}$. 
The negative sign between the first and second terms 
of \eq{Eq_2} indicates the electron motion from the QW to the cold lead.
The current carried by the electrons in the total system is calculated 
from the reduced density operator $\hat{\rho}_S^{h,c}$,
describing the state of the electrons in the QW under the influence of the 
reservoirs \cite{Vidar:ANDP201500298}.

In a steady state the left and right currents are of the same magnitude. 
We use a non-Markovian generalized master equation (GME) to describe the time-dependent 
electron motion in the system \cite{Vidar:ANDP201500298}.
The time needed to reach the steady state depends on the chemical potentials in each
reservoirs, the bias window, and their relation to the energy spectrum of the 
system \cite{ANDP:ANDP201600177}. 
In anticipation that the operation of an optoelectronic circuit can be sped up
by not waiting for the exact steady state we integrate the GME to $t=220$ ps, a point in
time late in the transient regime when the system is approaching the steady state.    

The QW is exposed to a uniform external perpendicular magnetic field
and is in a quantized photon cavity with a single photon mode. 
Therefore, we write the vector potential in the following form 
\begin{equation}
 \mathbf{A}(\mathbf{r}) = \mathbf{A}_{B}(\mathbf{r}) +\mathbf{A}_{\gamma}(\mathbf{r}),
\end{equation}
where $\mathbf{A}_{B}(\mathbf{r}) = -By \hat{\mb{x}}$ is the vector potential 
of the external magnetic field defined in the Landau gauge, and 
$\hat{\mathbf{A}}_{\gamma}$ is the vector potential of the photon field given by 
$\hat{\mathbf{A}}_{\gamma}(\mathbf{r})=A(\hat{a}+\hat{a}^{\dagger}) \mathbf{e}$. Herein,
the amplitude of the photon field is defined by $A$ with the electron-photon coupling constant 
$g_{\gamma}=eA a_w \Omega_w/c$,
and $\hat{a}$($\hat{a}^{\dagger}$) are annihilation(creation) operators 
of the photon in the cavity, respectively. 
The parameter that determines the photon polarization is $\mathbf{e}$ with either 
parallel polarized photon field $\mathbf{e} = \mathbf{e}_x$ or 
perpendicular polarized photon field $\mathbf{e} = \mathbf{e}_y$, 
The effective confinement frequency is $\Omega_w = \sqrt{\Omega^2_0 + \omega^2_c}$ with 
$\Omega_0$ being electron confinement frequency
due to the lateral parabolic potential and $\omega_c$ the cyclotron frequency due to the 
external magnetic field.

The QW is hard-wall confined in the $x$-direction and parabolically confined in
the $y$-direction. The Hamiltonian of the QW coupled to a single photon mode 
in an external perpendicular magnetic field in the $z$-direction is
\begin{align}\label{H_S}
 \hat{H}_{S}&=\int d^2r\;\hat{\psi}^{\dagger}(\mathbf{r}) \left[\frac{1}{2m^{*}}\left(\frac{\hbar}{i}\nabla+
              \frac{e}{c}   \mathbf{A}(\mathbf{r})  \right)^2 \right] \hat{\psi}(\mathbf{r})  \nonumber \\
            & +  \hat{H}_{ee} +\hbar \omega_{\gamma} \hat{a}^{\dagger}\hat{a}.
\end{align}
where $\hat{\psi}$ is the electron field operator. The second term of \eq{H_S} ($\hat{H}_{ee}$)  
represents the Coulomb electron-electron interaction in the quantum wire \cite{Nzar_IEEE_2016,Nzar.25.465302} and
the last term is the quantized single-mode photon field, with photon energy $\hbar\omega_{\gamma}$.
The electron-electron and the electron-photon interactions are taken into account stepwise using 
exact diagonalization techniques and truncations \cite{ABDULLAH2016280, Vidar:ANDP201500298}.
The time evolution of the system is described by a non-Markovian generalized master equation in order to study 
the non-equilibrium electron transport in the total system \cite{PhysRevB.85.075306}.

\section{Results}\label{Results}

In this section, we present results of our study of the thermoelectric transport. 
The system is a two-dimensional quantum wire in $xy$-plane with hard-wall confinement in the $x$-direction 
and parabolically confinement in the $y$-direction.  The quantum wire is connected to two leads with different 
temperature and the same chemical potential. We assume the temperature of the left lead ($T_{\rm h}$) is higher than 
that of the right lead ($T_{\rm c}$). The electron confinement energy of the quantum wire is equal to that of 
the leads $\hbar \Omega_w = \hbar \Omega_l = 2.0$~meV. The total system is exposed to an external low magnetic field 
$B = 0.1$~T.
The quantum wire is also coupled to a photon cavity with a single photon mode and 
the photons are linearly polarized in direction of electron propagation in the quantum wire ($x$-direction). 

\begin{figure}[htb]
\centering
    \includegraphics[width=0.26\textwidth,angle=0]{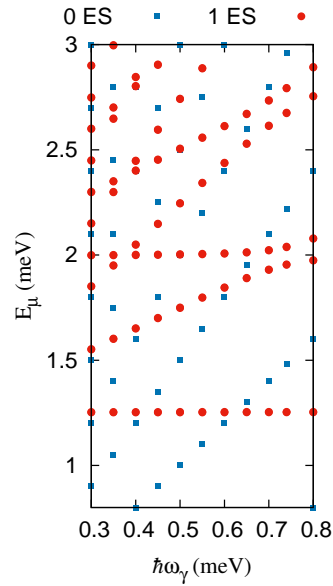}
 \caption{(Color online) Energy spectrum $E_{\mu}$ of the quantum wire versus the photon energy ($\hbar \omega_{\gamma}$), 
         0ES (blue rectangles) are zero-electron states, and 1ES (red circles) are one-electron states. 
         The electron-photon coupling strength is $g_{\gamma} = 0.05$~meV and
         the photons are linearly polarized in the $x$-direction. 
         The magnetic field is $B = 0.1~{\rm T}$, and $\hbar \Omega_0 = 2.0~{\rm meV}$.}
\label{fig01}
\end{figure}

Figure \ref{fig01} shows the energy spectrum ($E_{\mu}$) of the quantum wire versus the photon energy ($\hbar \omega_{\gamma}$),
$\rm 0 ES$ are zero electron states (blue rectangles) and $\rm 1 ES$ are the one-electron states (red circles). 
The electron-photon coupling strength is $g_{\gamma} = 0.05$~meV.
The state at $E_{\mu} = 1.25$~meV is the one-electron ground-state and the state at 
$E_{\mu} = 1.99$~meV is the first-excited state (red circles).
In addition, the state between the ground state and the first-excited state is the one-photon replica of the 
ground state that is located between $1.5\text{-}2.0$~meV for the selected range of the photon energy.
The one photon replica state at $\hbar \omega_{\gamma} = 0.3$~meV is located at $\sim 1.55$~meV while the same state 
at $\hbar \omega_{\gamma} = 0.74$~meV is seen at $\sim 1.99$~meV.
Since the photon energy at $0.74$~meV is approximately equal to the energy spacing 
between the ground-state and the first-excited state, the electronics system is then close to a 
resonance with the photon field.

\begin{figure}[htb]
    \includegraphics[width=0.5\textwidth,angle=0]{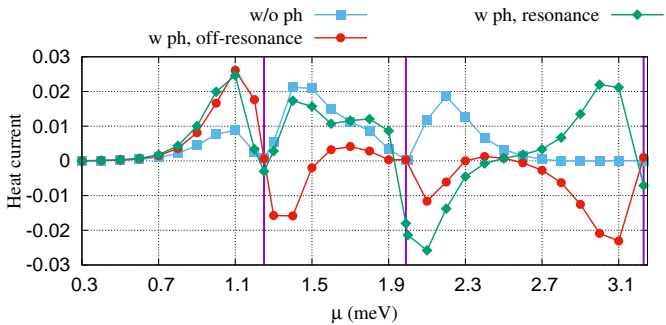}
 \caption{(Color online) Heat current as a function of the chemical potential 
                         $\mu = \mu_L = \mu_R$ plotted at time $t = 220$~ps
                         without a photon (w/o ph) cavity (blue squares), and with a photon cavity for the case of off-resonance (red circles), and 
                         resonance (green diamonds) at time $t = 220$~ps.
                         The temperature of the left (right) lead is fixed at $T_{\rm h} = 6.38$~K ($T_{\rm c} = 0.58$~K) implying 
                         thermal energy $k_B T_{\rm h} = 0.55$~meV ($k_B T_{\rm c} = 0.05$~meV), respectively. 
                         Therefore, the thermal energy of the system  $k_B \Delta T = k_B T_{\rm h} - k_B T_{\rm c} = 0.5$~meV.
                         The vertical lines (violet lines) display the location of a resonance 
                         of the leads with the three lowest states including the ground state, the first-excited state and the second-excited state at 
                         $E_{\mu} = 1.25$, $1.99$ and $3.23$~meV, respectively.
                         The photons are linearly polarized in the $x$-direction and the electron-photon coupling strength $g_{\gamma} = 0.15$.
                         The photon cavity is assumed to have one photon initially $N_{\gamma} = 1$. 
                         The magnetic field is $B = 0.1~{\rm T}$, and $\hbar \Omega_0 = 2.0~{\rm meV}$.}
\label{fig01_1}
\end{figure}

To demonstrate the transport properties we first present in \fig{fig01_1} the heat current versus the chemical potential 
$\mu = \mu_L = \mu_R$ without a photon (w/o ph) cavity (blue squares), and with a photon cavity for the off-resonance case (red circles), and 
resonance (green diamonds) at time $t = 220$~ps. The HC is plotted for 
the three lowest energy states at $E_0 = 1.25$, $E_1 = 1.99$ and $E_2 = 3.23$~meV, respectively.  
The vertical lines (violet lines) display the location of a resonance of the leads with the aforementioned three states.
In the absence of the photon field, the HC is zero at the resonance energy states and has positive value between the resonance energy states.
The positivity in HC can be explained by \eq{Eq_2_0}. Below the energy of each resonant state, meaning that the chemical potential 
is less than the energy of the state ($\mu < E_{\alpha}$), 
the first part of \eq{Eq_2_0},  ($\hat{\alpha} | \dot{\hat{\rho}}_S | \hat{\beta}$), is positive and the second part of equation,
($E_{\alpha} - \mu$), is also positive. Therefore, a positive value of HC is obtained. But, for energy above a resonant state 
when ($\mu > E_{\alpha}$) the first and the second part of the \eq{Eq_2_0} are both negative giving a positive value of HC. 
In addition, the HC is zero at resonance state because $E_{\alpha} = \mu$.

When we consider the QW system coupled to a cavity-photon field, the HC gives us a different interesting picture. 
The HC can assume negative values due to the dressed electron-photon states in both the off-resonance ($\hbar\omega_{\gamma} = 0.3$~meV)
and the resonance ($\hbar\omega_{\gamma} = 0.74$~meV) regimes.
In these cases, photon replica states participate to the transport. For instance, above the first-exited state at the energy $2.1$~meV, 
the first photon replica of the first-exited state participates to the transport leading to the negative HC.

\begin{figure}[htb]
  \includegraphics[width=0.5\textwidth,angle=0]{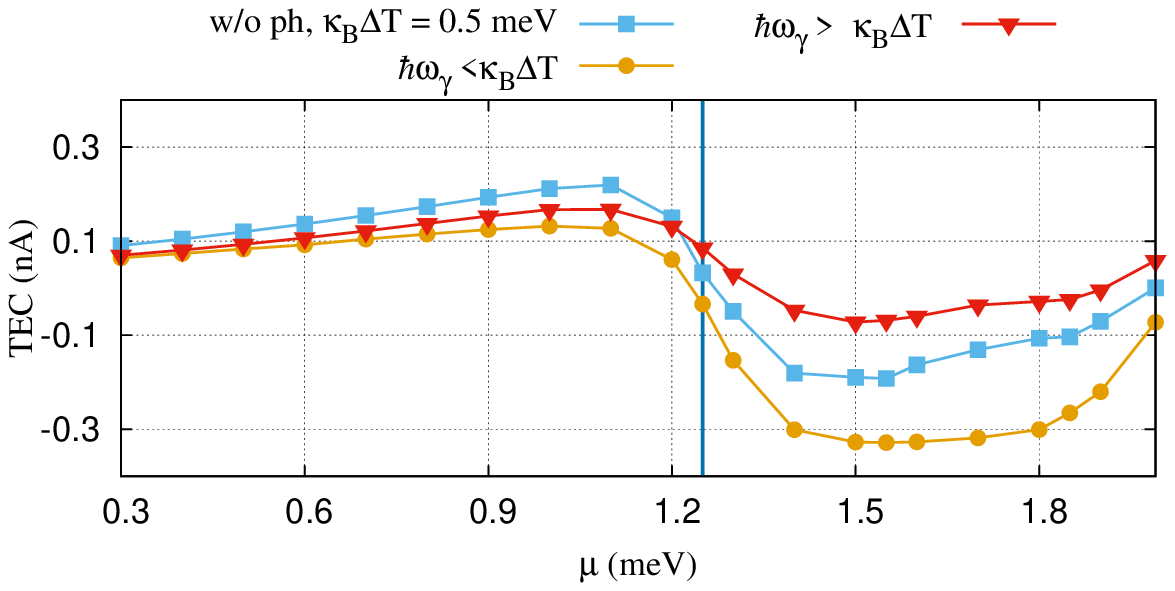}
 \caption{(Color online) TEC as a function of the chemical potential 
                         $\mu = \mu_L = \mu_R$ plotted at time $t = 220$~ps
                         without photon cavity (blue squares), 
                         and with photon cavity for the case of $\hbar\omega_{\gamma} < k_{\rm B} \Delta {\rm T}$ 
                         with $\hbar\omega_{\gamma} = 0.3$~meV (golden circles), 
                         $\hbar\omega_{\gamma} > k_{\rm B} \Delta {\rm T}$  with $\hbar\omega_{\gamma} = 0.74$~meV (red triangles). 
                         The temperature of the left (right) lead is fixed at $T_{\rm h} = 6.38$~K ($T_{\rm c} = 0.58$~K) implying 
                         thermal energy $k_B T_{\rm h} = 0.55$~meV ($k_B T_{\rm c} = 0.05$~meV), respectively. 
                         Therefore, the thermal energy of the system  $k_B \Delta T = k_B T_{\rm h} - k_B T_{\rm c} = 0.5$~meV.
                         The vertical line (blue line) displays the location a resonance 
                         of the leads with the ground state.
                         The photons are linearly polarized in the $x$-direction and the electron-photon coupling strength $g_{\gamma} = 0.15$.
                         The magnetic field is $B = 0.1~{\rm T}$, and $\hbar \Omega_0 = 2.0~{\rm meV}$.} 
                         
\label{fig02}
\end{figure}

Figure \ref{fig02} shows 
the thermoelectric current (TEC) versus the chemical potential $\mu = \mu_L = \mu_R$ 
without a photon (w/o ph) cavity (blue squares), 
and with a photon cavity for the case of $\hbar\omega_{\gamma} < k_{\rm B} \Delta {\rm T}$ (golden circles), and 
$\hbar\omega_{\gamma} > k_{\rm B} \Delta {\rm T}$ (red triangles) at time $t = 220$~ps. 
At this time point, the system is in the late transient regime close to a steady state.  

The current is essentially governed by the difference between the two Fermi functions
of the external leads. 
The thermoelectric current is generated when the Fermi functions of the leads have 
the same chemical potential but different step width. 
One can explain the TEC of the system without the photon cavity as the following: 
The TEC becomes zero in two cases. First, when the two Fermi functions are equal to $0.5$ (half filling) and in the second one 
both Fermi functions are $0$ or $1$ (integer filling)~\cite{Tagani201336, PhysicaE.53.178}.
Therefore, the TEC is approximately zero at $\mu = 1.25$~meV (blue squares) in the case 
of the system without the photon cavity corresponding to half filling of the ground state~\cite{Nzar_ACS2016}. 
The TEC is approximately zero at $\mu = 0.1$ and $1.99$~meV for the integer filling of occupation 
0 and 1, respectively.

We consider the quantum wire coupled to a photon field with initially one photon in 
the cavity in the $x$-polarization. We assume two different regimes. First, when the 
photon energy is smaller than the thermal energy ($\hbar\omega_{\gamma} < k_{\rm B} \Delta {\rm T}$)
in which the photon and the thermal energies are assumed to be $0.3$ and $0.5$~meV, respectively.
In this case the TEC is suppressed in ``positive'' and enhanced in ``negative'' part as is 
shown in \fig{fig02} (golden circles). The enhancement of TEC here is due to participation of 
one photon replica of the ground state around $E_{\mu} = 1.55$~meV at  $\hbar\omega_{\gamma} = 0.3$~meV (\fig{fig01}) 
to the electron transport instead of the ground state itself.
Generally, we expect no change in the TEC in the presence of a photon field here because 
the thermal energy is higher than the photon energy, i.e.\ the thermal smearing could be expected to prevent 
changes in the TEC. But, the one photon replica state strongly participates in the transport leading to 
an increase in the TEC. 

This inspires us to try the opposite:
We increase the photon energy to $0.74$~meV and keep the thermal energy at $0.5$~meV.
The photon energy is now approximately equal to the 
energy spacing between the ground state and the first-excited state of the the quantum wire as is clearly seen in 
\fig{fig01} at $\hbar \omega_{\gamma} = 0.74$~meV.
Therefore, the electronic system is almost in a resonance with the photon cavity, but under the condition of $\hbar\omega_{\gamma} > k_{\rm B} \Delta {\rm T}$ we now see in (\fig{fig02} - red triangles) that the TEC is
suppressed in both the ``positive'' and ``negative'' values due to the Rabi splitting.
In this case, the following states contribute to the transport: The ground state, the one- and two-photon replica 
states of the ground state, and 
the second excited state. The contribution of these states marks the presence of Rabi oscillations. 
Consequently, suppression of TEC in the system is observed~\cite{Nzar_ACS2016}.

\begin{figure}[htb]
  \includegraphics[width=0.5\textwidth,angle=0,bb=50 95 410 255]{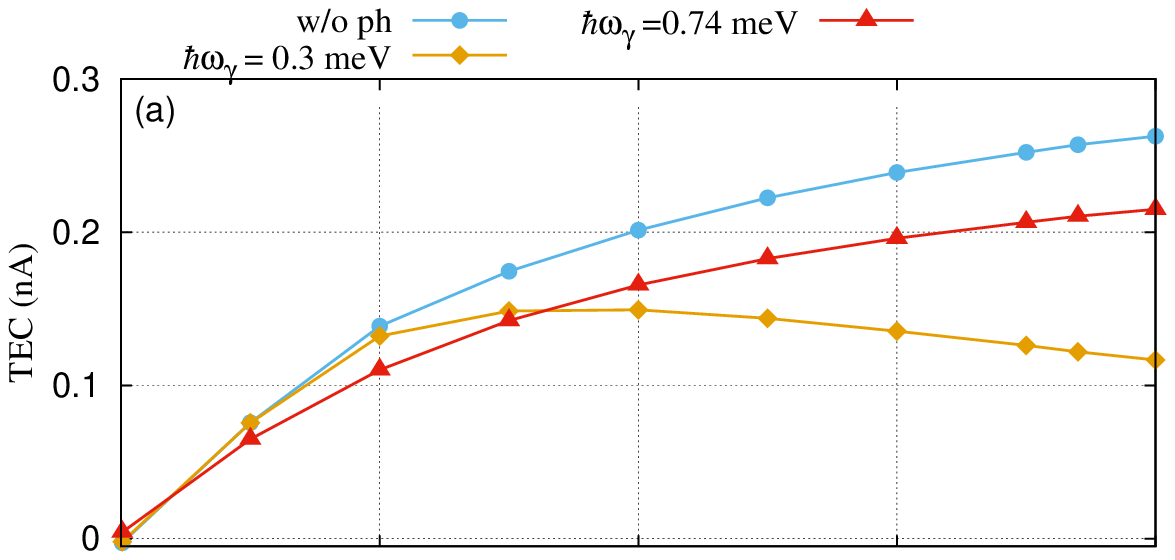}\\
  \includegraphics[width=0.5\textwidth]{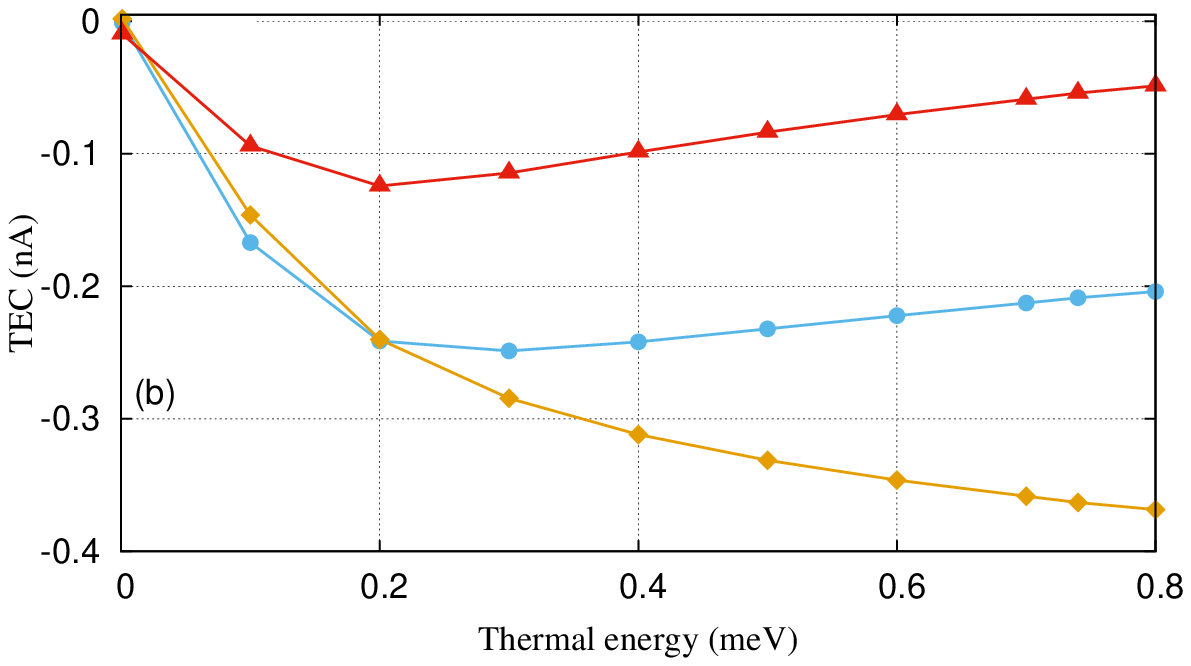}
 \caption{(Color online) TEC as a function of the chemical potential 
                         $\mu = \mu_L = \mu_R$ at time $t = 220$~ps is plotted 
                         below the GS at $\mu = 1.1$~meV (a) and above the GS at $\mu = 1.4$~meV (b)
                         shown in \fig{fig01} in the case of the system without the cavity (blue circles), 
                         and with the cavity for the photon energy $\hbar\omega_{\gamma} = 0.3$~meV (golden diamonds), 
                         and $\hbar\omega_{\gamma} = 0.74$ (red triangles). 
                         The photon is linearly polarized in the $x$-direction and the electron-photon coupling strength $g_{\gamma} = 0.15$ meV.
                         The temperature of the right lead is fixed at $T_{\rm c} = 0.58$~K implying 
                         thermal energy $k_B T_{\rm c} = 0.05$~meV, and the temperature of the left lead is tuned.
                         The magnetic field is $B = 0.1~{\rm T}$, and $\hbar \Omega_0 = 2.0~{\rm meV}$.}
\label{fig03}
\end{figure}

Now, we tune the thermal energy and keep the photon energy. 
The TEC versus thermal energy without photon cavity (blue circles), 
and with photon cavity for the case of $\hbar\omega_{\gamma}= 0.3$~meV (golden diamonds), and 
$\hbar\omega_{\gamma} = 0.74$~meV (red triangles) at $t = 220$~ps is shown in \fig{fig03}.

Figure \ref{fig03}(a) shows the TEC below the ground state at $\mu = 1.1$~meV and \fig{fig03}(b) 
presents the TEC above the ground state at $\mu = 1.4$~meV.
In the presence of the photon field, we see the same trends for the properties of the TEC mentioned above 
occurring for different values of thermal energies. 
For instance, if the photon energy is $\hbar\omega_{\gamma}  = 0.3$~meV (golden diamonds)
the TEC is almost unchanged up to $k_{\rm B} \Delta {\rm T} = 0.2$~meV. In addition, the rate of 
reduction of the TEC at $\mu = 1.1$~meV 
is almost equal to the rate of increasing of the TEC at $\mu = 1.4$~meV after $0.3$~meV.
For photon energy $0.74$~meV (red triangles),
the rate of increasing of the TEC at $\mu = 1.1$~meV is 
slower than the rate of reduction of the TEC at $\mu = 1.4$~meV.
As we have mentioned above, in the resonance photon energy regime ( $0.74$~meV) several states 
contribute to the electron transport such as
the ground state, the one- and two-photon replica states of the ground state, and 
the second excited state.
At high thermal energy, the charging of few of these states such as the ground state is getting weaker. 
This leads to a stronger influence of the Rabi oscillation between other contributing states. 
Therefore, the TEC at $\mu = 1.4$~meV is more suppressed around $\hbar\omega_{\gamma} = 0.74$~meV.

\begin{figure}[htb]
\centering
\includegraphics[width=0.25\textwidth,angle=0,bb=90 50 240 295]{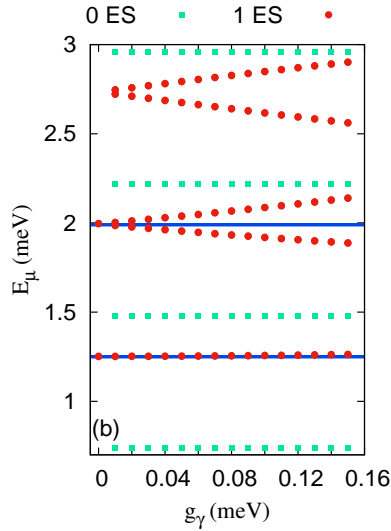}
 \caption{(Color online) Energy spectrum of the quantum wire versus electron-photon coupling strength $g_{\gamma}$.
         0ES (golden circles) are zero-electron states, 1ES (blue squares) are one-electron states 
         and the horizontal lines (red lines) display the location of the resonances 
         of the leads with the three lowest one-electron states in the case of the resonance photon field ($\hbar\omega_{\gamma} = 0.74$~meV).
         The photons are linearly polarized 
         in the $x$-direction. The magnetic field is $B = 0.1~{\rm T}$, and $\hbar \Omega_0 = 2.0~{\rm meV}$.}
\label{fig04}
\end{figure}

In order to show further the influence of the photon field on TEC, we tune the electron-photon coupling strength. 
Figure \ref{fig04} demonstrates the energy spectrum $E_{\mu}$ versus the electron-photon coupling strength $g_{\gamma}$, 
0ES  (green squares) are zero-electron states, 1ES (red circles) are one-electron states, and 
the horizontal lines (blue dotted lines) show the location of the the two lowest one-electron states in
resonance with the leads with. The photon energy is $0.74$~meV.

We begin with vanishing electron-photon coupling strength $g_{\gamma} = 0$~meV.
Two one-electron states are observed in the selected range of the energy spectrum:
The ground state and the first-excited state with energy values $E_0 = 1.25$~meV and 
$E_1 = 1.99$~meV, respectively. In the presence of the cavity with photon energy 
$\hbar \omega_{\gamma} \simeq E_1 - E_0 \simeq 0.74$~meV
the one-photon replica of the ground state is formed near to the first-excited state indicating 
a resonance regime.
Increasing the electron-photon coupling strength, more splitting between the one photon replicas
of the ground state and the first-excited state are seen. The splitting here is 
the Rabi splitting. We should mention that the same splitting occurs between the two-photon replica 
of the ground state and the one-photon replica of the first-excited state between $2.5\text{-}3.0$~meV.
Even though we refer to photon replicas here with a certain photon number the reader should have in
mind that photon replicas are a perturbational view of dressed electron states, that close to a resonance
do not have an integer number of photons associate with them. The dressed states generally include a 
linear combination of several of the eigenstates of the photon number operator.

\begin{figure}[htb]
  \includegraphics[width=0.5\textwidth,angle=0,bb=50 95 410 255]{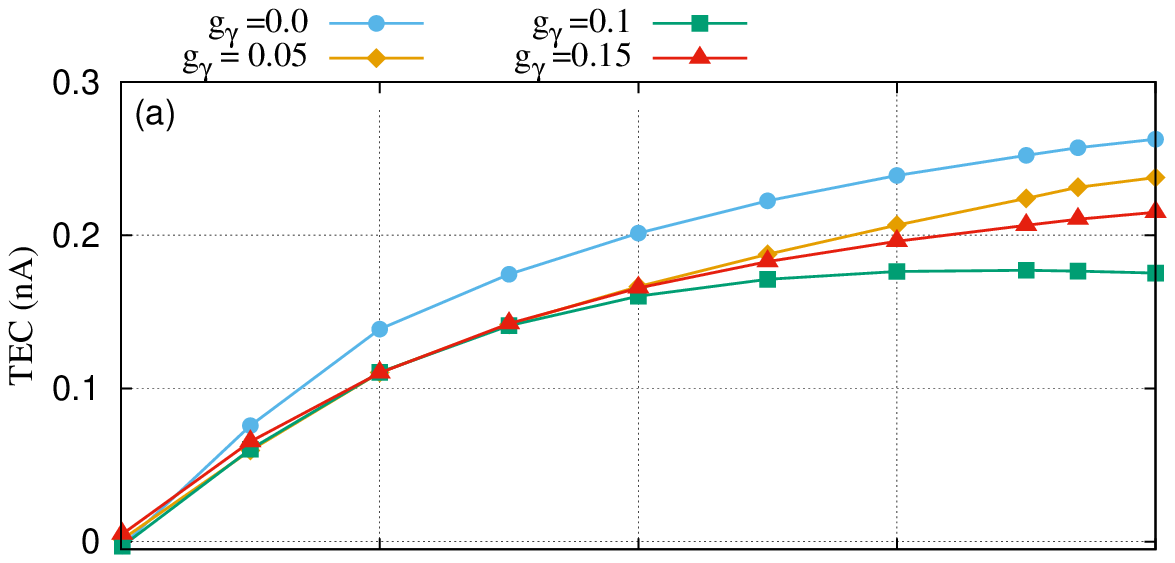}\\
  \includegraphics[width=0.5\textwidth]{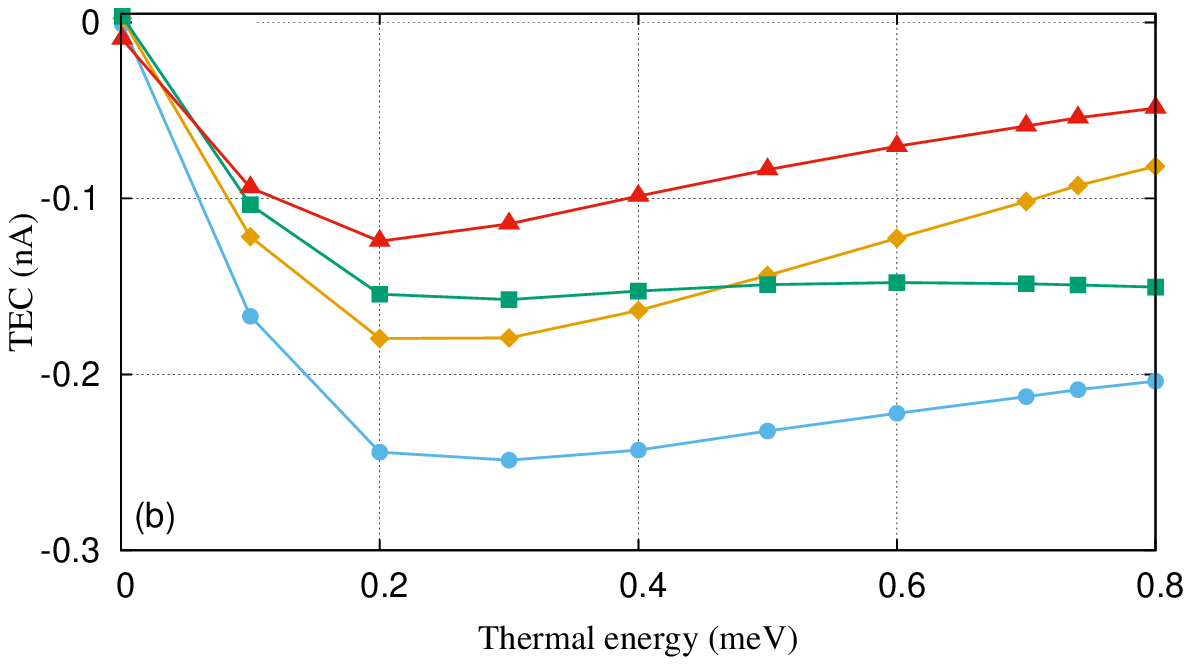}
 \caption{(Color online) TEC as a function of the thermal energy ($k_{\rm B} \Delta T$) 
                         at time $t = 220$~p is plotted below ($\mu = 1.1$~meV) (a) 
                         and above ($\mu = 1.4$~meV) (b) the ground state
                         shown in \fig{fig01}                     
                         for the system without the cavity $g_{\gamma} = 0.0$~meV (blue circles), 
                         and with the cavity with the electron-photon coupling strength 
                         $g_{\gamma} = 0.05$ (golden diamonds), $0.10$ (green squares), 
                         and $0.15$~meV (red triangles). The photon energy is
                         $\hbar \omega_{\gamma} \simeq E_1 - E_1 \simeq 0.74$~meV (resonance regime)
                         and the photons are linearly polarized in the $x$-direction.
                         The temperature of the right lead is fixed at $T_{\rm c} = 0.58$~K implying 
                         thermal energy $k_B T_{\rm c} = 0.05$~meV, and the temperature of the left lead is tuned.
                         The magnetic field is $B = 0.1~{\rm T}$, and $\hbar \Omega_0 = 2.0~{\rm meV}$.}
\label{fig05}
\end{figure}
Figure \ref{fig05} shows the TEC as a function of the thermal energy at time $t = 220$~ps 
and $\mu = 1.1$~meV (a) and $\mu = 1.4$ (b) without 
a photon field $g_{\gamma} = 0.0$~meV (blue circles) and with photon field for the 
electron-photon coupling strength $g_{\gamma} = 0.05$ (golden diamonds), $0.1$ (green squares) 
and $0.15$~meV (red triangles) in the case of a nearly resonant photon field 
($E_1 - E_0 \simeq \hbar \omega_{\gamma} = 0.74$~meV).
We fix the temperature of the right lead at $T_c = 0.58$~K implying $k_{\rm B} T_c = 0.05$~meV and 
tune the temperature of the left lead. Clearly seen is a reduction in the TEC with
the electron-photon coupling strength. This reduction in the TEC is a direct consequence of the 
increased Rabi splitting shown in \fig{fig04}. 
In a previous publication, we demonstrated the effects of electron-photon coupling strength
on the TEC for a fixed value of thermal energy. We showed that Rabi splitting causes a reduction
in the TEC with increasing electron-photon coupling strength. The stronger electron-photon coupling 
strength, the more reduction of the TEC was seen~\cite{Nzar_ACS2016}. 
Here, we present the same influences of the electron-photon coupling but for different regimes of thermal energy.  
We show that it is not necessary to have a stronger reduction of the TEC for a stronger electron-photon coupling
at high thermal energy especially between $0.5\text{-}0.8$~meV.
In \fig{fig05}(b) it is seen that the TEC for $g_{\gamma} = 0.1$~meV is higher than that of $g_{\gamma} = 0.05$~meV between  
$0.5\text{-}0.8$~meV and the TEC for $g_{\gamma} = 0.15$~meV is higher than that of $g_{\gamma} = 0.1$~meV 
shown in \fig{fig05}(a).
So, it demonstrates that the effects of the electron-photon coupling strength may 
influenced by the thermal energy in the system. Therefore, 
the influence of the Rabi-oscillation is modified by higher thermal energy in the system.

\section{Conclusions}
\label{ConclusionS}

In this article we have shown photo-thermo-electric transport described by a quantum 
master equation, taking advantage of exact-diagonalization to include electron-photon interaction.
We demonstrate that the heat current can be controlled in unexpected ways by a cavity-photon field.
In addition, we show that the calculated thermoelectric current through the system depends on 
the ratio between the photon energy and the thermal energy, and the electron-photon coupling strength.

We observe that the effects of the cavity-photons on the thermoelectric current depend on the location
of the chemical potential with respect to a particular state of the central system, and the ratio of
the photon and the thermal energy. For low thermal energy the photons do enhance the thermoelectric current.
As expected the largest effects are close to a Rabi resonance,
when the electrons and photons are strongly coupled, and when the thermal energy is large.
In this case, with the chemical potential of the leads just above the transport-resonant state the cavity photons
reduce the transport current in the system. An off-resonant photon field enhances the 
transport current though. On the other hand, when the transport-resonant state is just below the chemical
potential of the leads, the cavity-photons reduce the current for both cases, for nonresonant and 
resonant photons, but stronger in the latter case.
The partially negative heat current is a result of a certain energy transfer from the photons to the electrons,
as the system has not completely reached a steady state. The photons create an external perturbation on the electrons 
which behave like a slightly driven subsystem~\cite{ABDULLAH2018102}.

As a consequence, the thermoelectric current can be controlled in nanostructures using a quantized cavity photon field 
which may benefit energy harvesting in devices made thereof. 

\section*{Acknowledgment}
This work was financially supported by the Research
Fund of the University of Iceland, the Icelandic Research
Fund, grant no. 163082-051. The calculations were carried out on 
the Nordic high Performance computer (Gardar). We acknowledge the University of Sulaimani, Iraq.

\section{References}

\bibliographystyle{elsarticle-num} 
%\bibliography{Ref.bib}

%
\end{document}